\documentclass[conference]{IEEEtran}
\IEEEoverridecommandlockouts
\usepackage{cite}
\usepackage{amsmath,amssymb,amsfonts}
\usepackage{algorithmic}
\usepackage{multirow, array, subcaption}
\usepackage{tikz}
\usepackage{xspace}
\usepackage{adjustbox}
\usepackage{graphicx}
\usepackage{textcomp}
\usepackage{xcolor}
\usepackage{hyperref}
\usepackage[flushleft]{threeparttable}

\newcommand{\unet}{U-Net\xspace}

\newcommand{\musdb}{\texttt{musdb18}\xspace}
\newcommand{\waveunet}{Wave-U-Net\xspace}

\usetikzlibrary{shapes.misc}

\newcommand{\croix}[1][]
{%
\begin{tikzpicture}[baseline=(textbox.base),inner sep=0pt]
\node[cross out,draw,text width=\dimexpr#1] (textbox) {\strut};
\useasboundingbox (textbox);
\end{tikzpicture}%
}
\renewcommand{\baselinestretch}{0.992}

\def\BibTeX{{\rm B\kern-.05em{\sc i\kern-.025em b}\kern-.08em
    T\kern-.1667em\lower.7ex\hbox{E}\kern-.125emX}}
\begin{document}

\title{Improving singing voice separation using Deep U-Net and Wave-U-Net with data augmentation
}

\author{\IEEEauthorblockN{Alice Cohen-Hadria}
\IEEEauthorblockA{\textit{UMR STMS 9912}\\
\textit{Sorbonne Universit\'e, IRCAM, CNRS,}\\
Paris, France \\
cohenhadria@ircam.fr}
\and
\IEEEauthorblockN{Axel Roebel}
\IEEEauthorblockA{\textit{UMR STMS 9912} \\
\textit{IRCAM, Sorbonne Universit\'e, CNRS,}\\
Paris, France \\
axel.roebel@ircam.fr}
\and
\IEEEauthorblockN{Geoffroy Peeters}
\IEEEauthorblockA{\textit{LTCI T\'el\'ecom ParisTech} \\
\textit{Universit\'e Paris-Saclay}\\
Paris, France \\
geoffroy.peeters@telecom-paristech.fr}
}

\maketitle

\begin{abstract}
State-of-the-art singing voice separation is based on deep learning making use of CNN structures with skip connections (like \unet~model, \waveunet~model, or  MSDENSELSTM). A key to the success of these models is the availability of a large amount of training data. In the following study, we are interested in singing voice separation for mono signals and will investigate into comparing the \unet and the \waveunet that are structurally similar, but work on different input representations. First, we report a few results on variations of the \unet model. Second, we will discuss the potential of state of the art speech and music transformation algorithms for augmentation of existing data sets and demonstrate that the effect of these augmentations depends on the signal representations used by the model. The results demonstrate a considerable improvement due to the augmentation for both models. But pitch transposition is the most effective augmentation strategy for the  \unet~model, while transposition, time stretching, and formant shifting have a much more balanced effect on the \waveunet~model. Finally, we compare the two models on the same dataset. 

\end{abstract}



%

\section{Introduction}
In the case of music, source separation aims at separating the various instruments (such as the singing voice, guitar, piano or drums) present in the mixture (the mix).
When the source of interest is the singing voice, various assumptions can be made to help the separation, such as assuming a source/filter production mechanism \cite{durrieur2010}, using the sparsity in frequency of the vocals in Robust Principal Component Analysis (rPCA) \cite{candes2011_acm}, assuming the non-repetition of the vocal parts over time  \cite{Rafii2011_ASM} or using Non Negative Matrix factorization \cite{vemvu2005_ismir}.
Those assumptions lead to a first set of approaches for singing voice separation which are unsupervised, called Blind Audio Source Separation.\\
Recently, because of the availability of new annotated training datasets, supervised approaches have taken the lead, especially using neural networks methods.
The current state-of-the-art, and winner of SiSEC 2018 \cite{sisec2018}, is a combination Long-Short Time Memory networks and Dense Convolutional Neural Networks, presented in \cite{Takahashi2018} and use stereo signals.

In the following we will discuss the problem of singing voice separation using  mono signals. In this context \cite{chandna2017_lva} relies on Convolutional Neural Networks (ConvNet). A more sophisticated version of these, the deep \unet~(also called \unet) architecture has been proposed in \cite{jansson2017_ismir}.
Both process the spectrogram. They, therefore, necessitate the frequency-to-time reconstruction of the audio signal, which potentially leads to artifacts.
For this reason, Stoller proposed in \cite{stoller2018_ismir} the \waveunet~model which directly processes and separates the audio signal.
The comparison of these two models is of interest, since they share most of architecture properties, while processing very different inputs (temporal audio signals and spectrogram).
The goal of this paper is to compare these two models and the implication of using either audio signals or spectrograms as input for source separation.
The \waveunet~has been trained on a rather limited dataset, consisting of the train part of the \musdb~dataset ($\approx$6h) and the CCMixter dataset \cite{stoller2018_ismir} ($\approx$3h).
For \unet, \cite{jansson2017_ismir}, the authors used a private dataset containing approximately 20,000 tracks, ($\approx$2 months).
In the following we discuss a strategy to produce a data-set of comparable size from the publicly available \musdb~dataset, by means  of using state of the art signal transformation algorithms to produce various.
Using this augmented dataset, we will compare trained under the same conditions and will compare the effect of the different data augmentation strategies for the  \unet~and  \waveunet~ models.
We also conducted an in depth analysis of variations of the \unet~architecture (no skip-connections, comparing ratio masking with direct estimation of the separated source).
\\
In section \ref{sec:models}, we review in detail the previous works on \unet~and \waveunet.
Section \ref{sec:dataAugm} presents data augmentation for the singing voice separation problem and how we created a large dataset to train and evaluate our models in a conjoint framework.
Section \ref{sec:expeResults} presents the different experiments we conducted to study and compare \waveunet~and \unet~models and the results of these.

\section{Models}
\label{sec:models}
\subsection{\unet~model}
The \unet~model has been originally proposed for biological cells segmentation in \cite{ronneberger2015}.
Recently, Jansson has proposed in \cite{jansson2017_ismir} to use it for singing voice separation.
This model follows an encoder-decoder scheme.
\begin{figure}
\centering
\includegraphics[scale=0.37]{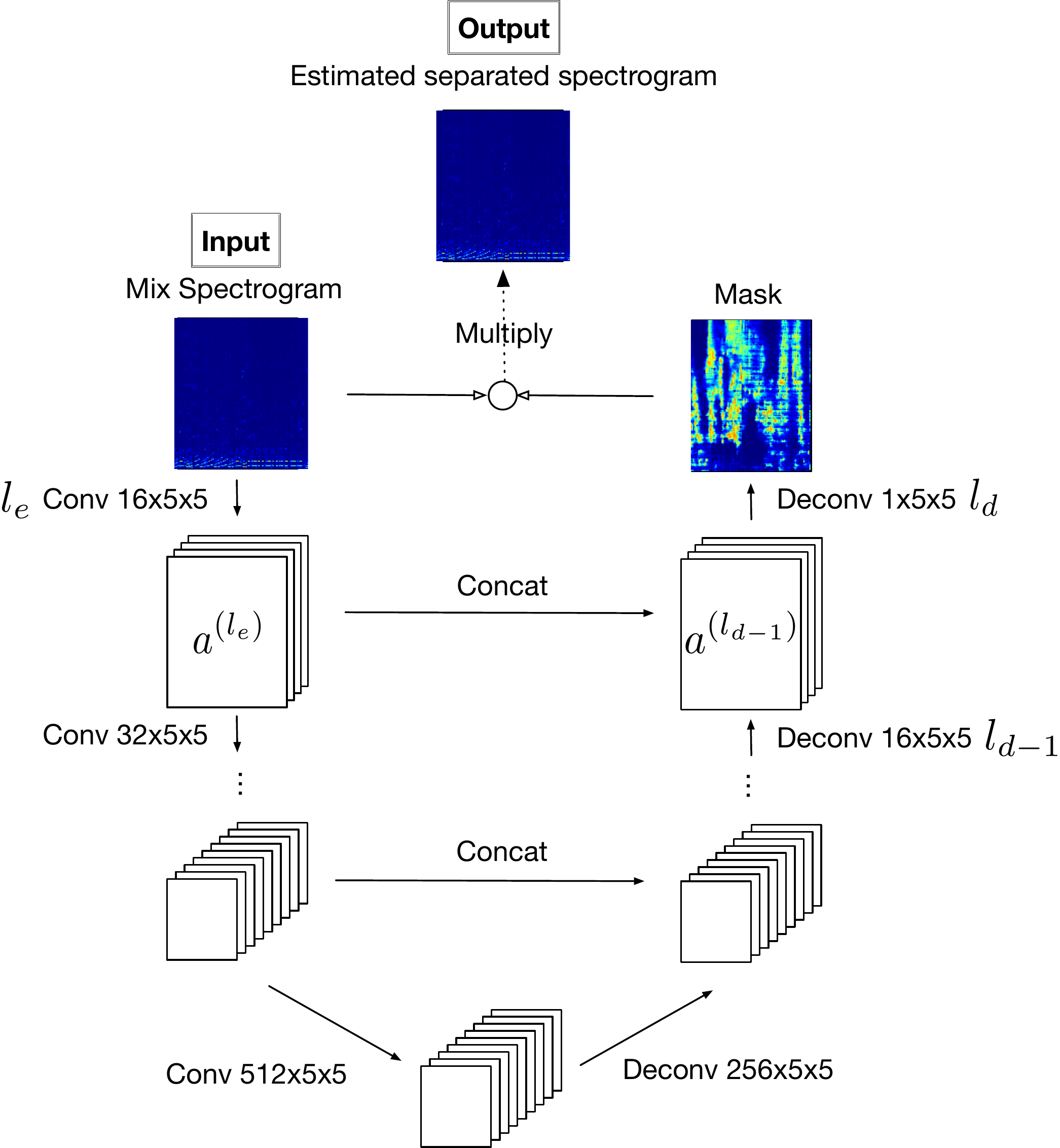}
\caption{\unet~architecture}
\label{fig:Unet}
\end{figure}
The \textbf{encoder} part of the network is made of a set of convolutional layers. 
The goal of the encoder is to reduce the inputs dimensionality while preserving relevant information for the task of interest.
The \textbf{decoder} part is made of deconvolutional layers. 
Usually, the decoder attempts to recreate the input from the compressed representation provided by the encoder. 
In this case, the model is called an Auto-Encoder (AE).
To apply this to singing voice separation, the input is a spectrogram excerpt of a mix track\footnote{For shorter notation, we will use spectrogram and spectrogram excerpt as synonyms in the remaining of this paper, and refer to full spectrogram for the spectrogram of the whole track.} and the output is a spectrogram of the isolated singing voice.
Compared to a classical AE, the \unet~model adds two novel ideas that will be tested in part \ref{sec:expeResults}. \\
\textbf{Skip connections.} 
Since the \unet~has a symmetric architecture (i.e. each layer couple $l_e$ in the encoder and  $l_d$ in the decoder have the same number of filters, sizes, strides and output dimensions), they can be connected through skip-connections (see figure \ref{fig:Unet}).
The motivating idea for these connections is to help the reconstruction by providing finer details in the decoding directly from the encoder (which are otherwise progressively lost during encoding). 
\\
\textbf{Output as mask.} 
In an usual AE, the decoder aims at reconstructing the given input.
In this specific case of source separation, the decoder aims at reconstructing the spectrogram of the isolated source. 
In the \unet~model, instead of defining the output as the spectrogram of the isolated source $Y$, the output is defined as a continuous (which values range from 0 to 1) mask $f_\theta(X)$ to be applied to the input spectrogram $X$ to obtain the spectrogram of the isolated source $Y$.
The loss function to be minimized is therefore defined as:
$
\mathcal{L (X,Y; \theta)} = \left\|f_\theta(X) \odot X - Y \right\|_1^1.
$
To test and analyze the specific features of the \unet, we conducted experiments (see \ref{sec:expeResults}) consisting in 1) outputting directly a separated spectrogram and 2) removing the skip connections.
\subsubsection{Details of the architecture, training and testing}
Each layer of the \unet~model is made of $5\times5$ filters with stride 2. 
The first layer has 16 filters and the number of filters is doubled at each layer.
The activation are Leaky ReLUs ($\alpha=0.2$).
A batch normalization layer is used between all layers.
%
The decoder part is mapped to the encoder part. The activations are ReLUs, expect for the last layer which uses sigmoid activations (to keep the values of the masks between 0 and 1).
Dropout ($p=0.5$) is applied on the first three layer of the decoder part.
Training is done using minibatch of size 128, and ADAM \cite{kingma20154} optimizer. \\
%
At test time, a track is processed by passing non-overlapping patches $X$ of 128 frames of the full spectrogram of the mix through the \unet. 
The full spectrogram of the isolated singing voice for the track is simply obtained by concatenating temporally the separated spectrogram outputed by the \unet, called $\hat{Y}$. 
The audio signal is reconstructed using $\hat{Y}$ and the phase of $X$.\\
%

\subsection{\waveunet~model}
A recent evolution of the U-Net is the Wave-U-Net \cite{stoller2018_ismir}, an
end-to-end network using a similar topology as the U-Net, but which works directly on the audio signal (therefore avoiding the problems related to reconstruction of the audio signal). 
The evaluation proposed in \cite{stoller2018_ismir} seems to indicate that the Wave-U-Net can achieve similar performance as the U-Net, but, due to the fact that for the evaluation in \cite{stoller2018_ismir} only a much smaller training database was available, the conclusions would benefit from an evaluation with the augmented dataset described in the following. \\
\textbf{Adaptation of \waveunet.} 
Due to the enormous size of the augmented dataset we are proposing, we adopt the strategy presented in \cite{jansson2017_ismir} and work with audio at 8192Hz sample rate. 
The change of the sample rate requires adaptation of the Wave-U-Net topology. 
From the many different possible choices, we chose to keep the time duration of the first layer 1-d filters approximately constant, reducing the filter size from 15 to 5 taps.
This filter length is the most similar time span under the constraints of the  Wave-U-Net. 
With respect to the receptive field of the model, the most similar setup compared to the evaluation in \cite{stoller2018_ismir} is given by a receptive field covering $\sim$7s (57431 samples).
This provides an output vector of about 1s (8197 samples). 
The corresponding values in \cite{stoller2018_ismir} are 6.7s (147443 samples, 22,05kHz) for prediction and $\sim$0.74s (16389 samples). \\
A particularity of \cite{stoller2018_ismir} is the fact that the training set is split into training and validation data, where the validation data is used to stop the training when no progress is made. 
For comparison with the original paper, we initially followed the same strategy, but later found that, with the augmented dataset, the problem of over-fitting is rather small.
Therefore, for the final comparison with U-Net, we used the full 100 samples of training data (with augmentation) in \musdb~for training. 
This example will be marked as DA-F in the results table. 
As initial experiment, and to confirm that our implementation for reduced sample rate performs correctly, we did use the optimal model for mono input (M3 in \cite{stoller2018_ismir}) and trained it on a random selection of 75 audio tracks without augmentation, using the other 25 tracks for early stopping after 20 epochs without improvement and evaluated this baseline model using the median SDR proposed as evaluation measure in \cite{stoller2018_ismir}. 
Training is done using minibatch of size 64, and ADAM \cite{kingma20154} optimizer. 
In our implementation with the slightly reduced filter size and slightly increased length of predicted output, we obtain a median SDR of 4.09dB while \cite{stoller2018_ismir} reports 3.96dB. Given the difference in network structure, the different split in training and validation data, the reduced sample rate, this value supports our idea that the network performs similar to the original in \cite{stoller2018_ismir}.\\ 
We present in the next section the methodology we implemented to create a very large dataset in order to train and compare \unet~and \waveunet. 

\begin{table*}[h!]
\caption{Datasets used}
\label{tab:dataset}
\centering
\begin{tabular}{c| c | c | c | c}
Acronym & training set & \#tracks for training & validation set & test set\\
\hline
no-DA &  75\% train set of \musdb~& 75 tracks & 25\% train set of \musdb& test set \musdb\\
DA &75\% train set of augmented \musdb~ &11 250 tracks & 25\% train set of augmented \musdb~ & test set \musdb\\
DA-F & 100 \%train set of augmented \musdb & 15 000 tracks& $\times$ & test set \musdb\\
\end{tabular}
\end{table*}
\section{Data augmentation for singing-voice}
\label{sec:dataAugm}
Data augmentation can be used to increase the number of training examples leading to an improved coverage of the real world signal space. To be able to augment training data without requirement for extensive re-annotation of the ground truth annotation (labels or separated signals), one needs to find means that modify the available training data such that the ground truth either does not change or changes in a predictable way so that it can be adapted automatically as part of the data augmentation procedure. In the following we focus the discussion on sound specific transformations (leaving aside transformations such as dropout or added noise). \\
Data augmentation of singing voice has been performed for singing voice  detection in \cite{schlueter2015_ismir}. In that case, the proposed transformations are applied directly on the  mel band spectrogram, treating it as an image. Accordingly, time stretching and pitch shifting the spectrogram is performed by means of dilated or compressed along the time - or frequency axis.\\ 
While these image transformations did improve results for the voice detection task, they seem less pertinent for source separation, where a precise link between waveform and spectrogram is of central importance. The operations used in \cite{schlueter2015_ismir} will change the form and width of the sinusoidal peaks and deform the attacks. At the end, the spectrogram does no longer represent any realizable signal. Moreover, we note that these spectrogram modifications can't be used for \waveunet, since it is not possible to retrieve the temporal signals once the magnitude spectrogram has been modified.\\
\cite{Uhlich2017_icassp} proposes rather basic strategies for data augmentation for singing voice separation: random swapping left/right channels for each instrument, random scaling with uniform amplitudes, random chunking into sequences for each instrument, and random mixing of instruments from different songs. The effect of the data augmentation evaluated on the DSD100 dataset remains rather limited, improving the results on average for 0.2dB for SDR metric and the vocal target on the test set of the DSD100 dataset [tab. 2]\cite{Uhlich2017_icassp}. We note that a few of the augmentation strategies rely on stereo data, a situation not covered by the present article. 

The software framework \texttt{muda} has been proposed in \cite{mcfee2015_ismir} as a flexible tool for augmenting musical datasets. The framework includes transformations comprising dynamic range  compression, mixing with noise, as well as time stretching and pitch shifting operations. The last two  operations are implemented using the open source library rubberband\footnote{\url{https://www.breakfastquay.com/rubberband/}} which according to its documentation is based on a phase vocoder algorithm that loosely implements the key points of state of the art phase vocoding: dedicated handling of transients (e.g. \cite{Roebel:03b}) and intra partial vertical phase coherence (e.g. \cite{Laroche/Dolson:99}). Shape-invariant processing, an essential feature for high quality speech-processing \cite{Roebel:10}, is not addressed, but might not be of major importance for approaches based on masking STFT magnitudes. More importantly, \texttt{muda} does not allow modifying the spectral envelope (formants) independently of the pitch, one of the key elements for voice transformation, avoiding for example the mickey mouse effect when transposing the pitch up.\\

\subsection{Proposed data augmentation strategy}
The data augmentation strategy used in the following experiments benefits from the fact that the \musdb~dataset
is provided in form of 4 separate signals containing: voice, drum, bass and accompaniment. 
Each of the four signals is transformed separately, selecting the musically and technically most appropriate signal processing parameters, as for example, excluding the drum signal from pitch shifting transformations.\\
The full set of transformations applied to the \musdb~tracks contains the following operations (transposition in cents):\\
- pitch-shifting but preserving the spectral envelop $\in [-300, -200, -100, 0, 100, 200, 300]$\\
- time-stretching $\in [0.5, 0.93, 1, 1.07, 1.15]$\\
- transformation of the spectral envelope only of the singing voice $\in [-150, -100, 0, 100,150]$\\
Combining  all these  modifications leads to 175 possible variants (including the original) of each track. 
Given that the \musdb~training data contains 100 tracks with a bit more than 6 hours, the augmented dataset (called DA) contains 15.000 tracks of music with a total duration of about 1.5 months of continuous music.\\
Specific considerations for the individual sources are as follows.
The singing voice is transformed by means of pitch shifting, formant shifting and time stretching using a state of the art shape invariant phase vocoder \cite{Roebel:10}. 
The formant shifting  is performed using the algorithm presented in \cite{Roebel/Rodet:05a}. 
The parameterization of the voice transformation algorithm is performed dynamically over time using as main control the $F_0$ calculated using the swipe $F_0$ estimation algorithm \cite{Camacho:07}. 
To achieve high quality formant shifting (or preservation), the order of the spectral envelope is adapted to the $F_0$ following  \cite{Roebel/Villavicencio/Rodet:07a}, such that the formant modification/preservation affects as good as possible the personality of the singing voice. 
The window size is adapted to be four times the local period. 
For the drum signal, only time stretching transformations are applied using the transient preservation algorithm described in \cite{Roebel:03b} and a fixed window size of 50ms. 
Finally, for the bass signal and the remaining accompaniments we apply a phase vocoder algorithm  \cite{Laroche/Dolson:99} again using transient preservation as in \cite{Roebel:03b}.
The transformations described above have been performed with the signal transformation kernel available in version 3 of the AudioSculpt program \cite{Bogaards2004} that can be scripted and controlled via the Unix command line.
\section{Experiments and results}
\label{sec:expeResults}

In the following experiments, we evaluate the separation using the mir\_eval-toolbox\footnote{\url{http://craffel.github.io/mir_eval/\#mir_eval.separation.bss_eval_sources}}.
We compute the following three metrics  \cite{Vincent2006_ieee}: Source-to-Interference Ratios (SIR),  Source-to-Artifact Ratios (SAR) and Source-to-Distortion Ratios (SDR) and report the median over the test database. As measure of variability, we use the median of the absolute deviation from the median (MAD)\cite{stoller2018_ismir}.

\begin{table*}[h!]
\setlength\tabcolsep{5pt}
	\caption{Results of experiments 1), 2) and 3)}
	\label{tab:resUnet}
	\begin{subtable}[t]{.5\textwidth}
		\caption{Results for the \waveunet models. Median and median absolute deviation (MAD) of SAR, SIR and SDR.}
		\label{tab:compUW}
		\begin{tabular}{| p{0.7cm}  p{2.4cm} p{0.6cm}  p{0.5cm}  p{0.6cm}  p{0.5cm}  p{0.6cm}  p{0.5cm}| }
			\hline
			Model & Dataset  & \multicolumn{2}{c}{SAR} & \multicolumn{2}{c}{SIR} & \multicolumn{2}{c|}{SDR} \\
			\cline{3-8}
			& used & med & MAD & med & MAD & med & MAD \\
			\hline
			\hline
			\multirow{4}{*}{W$^{8,2}$} 
			& no-DA &  5.52  & 1.96 & 10.87 & 2.02  & 4.09 & 2.07 \\
			& Stretch$^5$ &  5.60  & 1.86 & \textbf{12.11} & 2.99 & 4.20  & 1.58\\
			& Env.$^6$& 5.23 & 1.86 &11.22 & 2.32 & 3.77 & 1.61 \\
			&Pitch$^4$ & \textbf{6.09} & 1.65 & 10.68 & 2.40 & 4.18 & 1.99 \\
			& DA$^1$ & 5.86  & 1.63 & 12.02 & 2.19 & \textbf{4.67} & 1.71\\ 			\hline
			& DA-F$^1$ & \textbf{6.62}  & 1.80 &  \textbf{13.90} & 2.74 & \textbf{5.42} & 1.72\\
			\hline
			\hline
			W\cite{stoller2018_ismir} & No-DA+CCMix$^{11}$ &  \croix[5.4mm]  & \croix[5.4mm] &  \croix[5.4mm]  & \croix[5.2mm] & 3.96 & 3.0 \\[0.09cm]
			\hline
		\end{tabular} 
		
	\end{subtable}
	\hspace{0.15cm} 
	\begin{subtable}[t]{.4\textwidth}
		\caption{Results of \unet models. Median and MAD of SAR, SIR and SDR.}
		\label{tab:unet}
		
		\begin{tabular}  { | p{1cm}  p{1.2cm} p{0.6cm}  p{0.6cm}  p{0.6cm}  p{0.5cm}  p{0.6cm}  p{0.5cm} |}
			\hline
			Model & Dataset & \multicolumn{2}{c}{SAR} & \multicolumn{2}{c}{SIR} &\multicolumn{2}{c|}{SDR} \\
			\cline{3-8}
			& used & med & MAD & med & MAD & med & MAD\\
			\hline
			\hline
			\multirow{4}{*}{U$^7$} &no-DA-F$^{9}$ & 5.76 & 4.21 & 11.75 & 2.05 & 4.52 & 2.48\\
			& Stretch-F  &5.73 & 2.28 & 12.38 & 2.48 & 4.85 & 2.06 \\
			& Env.-F  & 6.06 & 2.28 & 11.06 & 2.66 & 4.55 & 2.24 \\
			&  Pitch-F & 6.35 & 2.21 & \textbf{12.69} & 2.69 &  \textbf{5.20} & 2.09\\
			& DA-F &  \textbf{6.40} & 2.20 & 11.98 & 2.37&\textbf{5.20} & 2.22 \\
			\hline
			\hline
			U \cite{jansson2017_ismir} & DS-priv$^{10}$ & \textbf{11.30} & \croix[5.2mm] & \textbf{15.31} & \croix[5.2mm] & \croix[5.2mm] & \croix[5.2mm] \\
			\hline
			\hline
				no-skip  &  DA-F & 5.60 & 2.39 & 9.92 & 2.08 & 3.44 & 2.13\\
			\hline
			no-mask  &  DA-F &4.87 & 3.25 &  \textbf{14.71 }& 3.50 & 4.18 & 3.27 \\ 
			\hline
		\end{tabular} 
	\end{subtable}
\begin{tablenotes}
     \small
     \item $^1$ DA: all Data Augmentation, $^2$ 25\% of \musdb kept for early stopping according to  \cite{stoller2018_ismir}, $^3$ MDB: \texttt{MedleyDB}, \cite{bittner2014_ismir}, for the testing phase,
        $^4$ Pitch:  pitch shifting, $^5$ Stretch: time stretching, $^6$ Env.: transformation of the singing voice spectral envelop,
       $^7$ U: \unet~model, $^8$ W: \waveunet~model, $^9$ F: Full \musdb, training on full training data - no early stopping, $^{10}$: private dataset see \cite{jansson2017_ismir}, $^{11}$ Musdb + CCMixter datasets\cite{stoller2018_ismir} M3 in table 2. 
    \end{tablenotes}
\end{table*}

For all experiments, the starting point is the \musdb~dataset \cite{Rafii2017}.
This dataset contains 150 tracks ($\sim$10h duration) of different styles.
The 150 tracks are split into 100 tracks for training, and 50 for testing. This dataset is called no-DA in Table \ref{tab:resUnet}.
Note that \cite{jansson2017_ismir} is evaluating the \unet~model on \texttt{MedleyDB} not using early stopping\footnote{MedleyDB is a dataset presented in \cite{bittner2014_ismir}. 46 (out of 122) tracks of \texttt{MedleyDB}  is actually included in \musdb.}. 
For the experiments with \waveunet \cite{stoller2018_ismir} used only 75\% of the training data of \musdb~to perform training and keeps 25\% data as a validation set used for early stopping. For comparison with \cite{stoller2018_ismir} we use early stopping with \waveunet and for comparison with \unet we train \waveunet on the full training data.  To distinguish these setups we  
denote experiments without early stopping with an F appended to the dataset, e.g. DA-F for the dataset with full augmentation and no hold out validation data.
See table \ref{tab:dataset} for a summary.\\
Like in \cite{jansson2017_ismir}, we use mono signals down-sampled at 8192 Hz to reduce storage space and training time.
For the \unet experiments, we use  STFTs with 1024 window length 1024 and and overlap 256.\\
\textbf{\waveunet model}.
Compared to  original results in \cite{stoller2018_ismir}, our adapted \waveunet, trained on no-DA (\musdb dataset without augmentation) performs slightly better: 4.09dB SDR 
where  \cite[M3 in table 2]{stoller2018_ismir} using an extended \musdb dataset reports 3.96dB.
An explanation might be that the evaluation in \cite{stoller2018_ismir} uses 22.05kHz sample rate while we use only 8192Hz.\\
Regarding the data augmentations, and using the early-stopping strategy used in \cite{stoller2018_ismir}, we can see that time stretching and pitch shifting alone have only minor impact, for the SDR (+0.1dB).
The transformation of the spectral envelope even has a negative impact:  from 4.09 dB (resp. 5.52) to 3.77 (resp. 5.23) for SDR (resp. SAR). Shifting only the envelop does not seem to provide useful augmentation. Still, using all augmentation strategies leads to a +0.6 db on SDR and a +1.15 db on SIR, which might indicate that pitch and formant transformation is necessary to provide useful augmentation.
The additional experiment  without early stopping (DA-F) yields another +0.7dB on SDR, and gives overall our best \waveunet-model. For such a large training database, early stopping doesn't seem to be beneficial.\\
\textbf{Unet model}
Like the \waveunet model, time stretching and envelop transposition do not have a strong impact.
The most effective transformation is pitch shifting, giving +1db in SDR (from 4.52 to 5.20).
Our hypothesis is that the other two augmentation have a minor effect on the variation in the spectral mask. Overall, using all the transformations proposed, the \unet model gave the best performance on the test set of \musdb, on both SAR (from 5.76dB to 6.40 dB) and SDR (from 4.52 to 5.20).
The results indicated in the first lines of Table \ref{tab:unet} have been obtained using the original \unet model, with skip connections and estimating masks $f_{theta}(X)$.
In order to further investigate those properties, we propose results for \unet trained without skip connections and outputting directly a spectrogram instead of a mask.
In line ``no-skip", we indicate the results obtained by only removing the skip-connections.
We see that it damages the results as they drop from 6.40 to 5.60 (for SAR), from 11.98 to 9.92 (SIR) and 5.20 to 3.44 (SDR).
This can be explained by the fact that, as expected, the skip connections bring a lot of details in the reconstruction, making the masks way sharper.
In line ``no-mask", we change the definition of the output: instead of estimating the masks $f_\theta(X)$ we directly estimate the spectrogram of the separated source $\hat{Y}$.
We see that it also damages the results as they drop from 6.40 to 4.87 (SAR) and 5.20 to 4.18 (SDR). \\
\textbf{Comparison \unet~and \waveunet.}
The two models are very close: they both follow the encoder-decoder paradigm and both use skip connections.
The difference between both is the input/output representation.
The \unet~processes spectrograms and hence necessitates an extra step to reconstruct the audio signal (necessary to evaluate the model and listen to the results) which is potentially prone to artifacts.
The \waveunet~processes directly  the temporal signals and hence does not necessitates any reconstruction.
However, the temporal signal is a lot more difficult to analyze. 
Comparing these two architectures is therefore quite interesting. Here we refer to results of dataset DA-F in Table \ref{tab:resUnet}.
We can see that the \waveunet model gives the best results for all metrics: 6.62db versus 6.40 db for SAR, 5.42 dB versus 5.20 db for SDR.
While this result seem to indicate an advantage for \waveunet,  we consider this for the moment as only a first element. The computational complexity of both networks needs to be taken into account and, given the large dataset, an increased complexity leading to improvements for both seems possible. Ongoing work will be reported in the future, including listening tests revealing the perceptual relevance of these quantitative measures.  

\vspace{-0.1cm}
\section{Conclusion}
\label{sec:ccl}
In this paper, we proposed a new set of data augmentations designed for singing voice detection.
We reviewed two singing voice separation state-of-the-art models: the \unet~model and the \waveunet~model.
With our data augmentation strategy, we produced a very large dataset, giving us a robust conjoint framework to compare these models.
We showed that the use of these augmentations improved the results over the \musdb~dataset, the largest publicly available dataset for singing voice separation, for both the \unet~and the \waveunet~model.
However for both models, the results are rather close, which is very interesting given the different representations taken as input by the two models. \\
We also studied the \unet~architecture. We proved that the skip connections of the model are crucial to reconstruct the singing voice separated spectrogram. 
We also showed that outputting masks rather than spectrograms yields better results.

\bibliographystyle{IEEEbib}
\bibliography{bibli.bib}

\end{document}